\documentclass[showpacs,amsmath,amssymb,prd,floatfix,twocolumn]{revtex4}

\usepackage{epsfig}
\usepackage{graphicx}
\usepackage{bm}
\usepackage{amsfonts}

\def\L{\Lambda}

\def\p{\partial}

\def\k{\kappa}

\newcommand{\xf}{\xi_{\phi}}
\newcommand{\xs}{\xi_{\sigma}}
\newcommand{\sff}{s_{\phi}}
\newcommand{\sss}{s_{\sigma}}
\newcommand{\s}{\sigma}

\newcommand{\be}{\begin{equation}}
\newcommand{\ee}{\end{equation}}
\newcommand{\bq}{\begin{eqnarray}}
\newcommand{\eq}{\end{eqnarray}}

\begin{document}

\title{Non-minimally coupled canonical, phantom and quintom models of holographic dark energy}

\author{M. R. Setare}
 \affiliation{Department of Science, Payame Noor University,\\
Bijar, Iran }

\author{E. N. Saridakis }
 \affiliation{Department of Physics,
University of Athens,\\ GR-15771 Athens, Greece}

\begin{abstract}
We investigate canonical, phantom and quintom models, with the
various fields being non-minimally coupled to gravity, in the
framework of holographic dark energy. We classify them and we
discuss their cosmological implications. In particular, we examine
the present value of the dark energy equation-of-state parameter
and the crossing through the phantom divide, and we extract the
conditions for a future cosmological singularity. The combined
scenarios are in agreement with observations and reveal
interesting cosmological behaviors.
\end{abstract}

 \maketitle

\section{Introduction}

Nowadays it is strongly believed that the universe is experiencing
an accelerated expansion, and this is supported by many
cosmological observations, such as SNe Ia \cite{1}, WMAP \cite{2},
SDSS \cite{3} and X-ray \cite{4}. These observations suggest hat
the universe is dominated by dark energy with negative pressure,
which provides the dynamical mechanism of the accelerating
expansion of the universe. Although the nature and origin of dark
energy could perhaps understood by a fundamental underlying theory
unknown up to now, physicists can still propose some paradigms to
describe it. In this direction we can consider theories of
modified gravity \cite{ordishov}, or field models of dark energy.
The field models that have been discussed widely in the literature
consider a cosmological constant \cite{cosmo}, a canonical scalar
field (quintessence) \cite{quint}, a phantom field, that is a
scalar field with a negative sign of the kinetic term
\cite{phant,phantBigRip}, or the combination of quintessence and
phantom in a unified model named quintom \cite{quintom}. The
quintom paradigm intends to describe the crossing of the
dark-energy equation-of-state parameter $w_\L$ through the phantom
divide $-1$ \cite{c14}, since in quintessence and phantom models
the perturbations could be unstable as $w_\L$ approaches it
\cite{c9}.

In addition, many theoretical studies are devoted to understand
and shed light on  dark energy, within the string theory
framework. The  Kachru-Kallosh-Linde-Trivedi model \cite{kklt} is
a typical example, which tries to construct metastable de Sitter
vacua in the light of type IIB string theory. Despite the lack of
a quantum theory of gravity, we can still make some attempts to
probe the nature of dark energy according to some principles of
quantum gravity. An interesting attempt in this direction is the
so-called ``holographic dark energy'' proposal
\cite{Cohen:1998zx,Hsu:2004ri,Li:2004rb,holoext}. Such a paradigm
has been constructed in the light of holographic principle of
quantum gravity  \cite{holoprin}, and thus it presents some
interesting features of an underlying theory of dark energy.
Furthermore, it may simultaneously provide a solution to the
coincidence problem, i.e why matter and dark energy densities are
comparable today although they obey completely different equations
of motion \cite{Li:2004rb}. The holographic dark energy model has
been extended to include the spatial curvature contribution
\cite{nonflat} and it has been generalized in the braneworld
framework \cite{bulkhol}. Lastly, it has been tested and
constrained by various astronomical observations \cite{obs3}.

In the present work we are interested in investigating various
field models of dark energy, where the fields are non-minimally
coupled to gravity \cite{Uzan:1999ch,nonminimal}. Such models have
been shown to present significant features and we study them in
the framework of holographic dark energy. In particular, we
examine the current value of $w_\L$ and the realization of a
recent crossing through the phantom divide $-1$ from above.
Additionally, we investigate the possibility of a future
$w_\L$-divergence \cite{phantBigRip,BigRip}, and the specification
of the time that is it going to happen. The plan of the work is as
follows: In section \ref{HDE} we construct the cosmological
scenarios of non-minimally coupled canonical, phantom and quintom
fields, in the framework of holographic dark energy. In section
\ref{cosmimpl} we examine their behavior and we discuss their
cosmological implications. Finally, in section \ref{conclusions}
we summarize our results.

\section{Non-minimally coupled
fields in the framework of holographic dark energy} \label{HDE}

Let us describe briefly the holographic dark energy framework
\cite{Cohen:1998zx,Hsu:2004ri,Li:2004rb,holoext}. In this
dark-energy model one determines an appropriate quantity to serve
as an infrared cut-off for the theory, and imposes the constraint
that the total vacuum energy in the corresponding maximum volume
must not be greater than the mass of a black hole of the same
size. By saturating the inequality one identifies the acquired
vacuum energy as holographic dark energy:
\begin{equation}
\rho_\Lambda=\frac{3c^2}{8\pi G L^2},\label{HDEdef}
\end{equation}
with $L$ the IR cut-off and $c$ a constant which can be set to 1.
Although the choice of $L$ has raised a discussion in the
literature \cite{Li:2004rb,Guberina}, in this work we will use the
Hubble scale. Note that the aforementioned choice for the IR
cut-off  has been found to have problems in conventional,
minimally-coupled frameworks \cite{Hsu:2004ri}, but this is not
anymore the case if one considers non-minimal coupling, as we do
in the present work. Finally, we mention that the extension of
holographic dark energy in the presence of non-minimally coupled
fields could possibly raise some theoretical questions, but we
assume that such an extension is valid. The detailed examination
of this subject is left for a future work.

In the following, we are going to investigate holographic dark
energy in the presence of canonical, phantom, or both fields,
non-minimally coupled to gravity. The space-time geometry will for
simplicity be a flat Robertson-Walker:
\begin{equation}\label{met}
ds^{2}=-dt^{2}+a(t)^{2}(dr^{2}+r^{2}d\Omega^{2}),
\end{equation}
with $a(t)$ the scale factor.

\subsection{Canonical field} \label{scalar}

We first consider a canonical scalar field with a non-minimal
coupling. This case has been partially investigated in \cite{Ito},
and here we extend it. The action of the universe is
\begin{equation}
S=\int d^{4}x \sqrt{-g} \left[\frac{1}{2\k^{2}}
R-\frac{1}{2}\,\xf\phi^{2} R
-\frac{1}{2}g^{\mu\nu}\p_{\mu}\phi\p_{\nu}\phi
+\cal{L}_\text{M}\right] \label{actioncan},
\end{equation}
where $\k^{2}$ is a gravitational constant. In the action we have
added a  canonical scalar field $\phi$, which in non-minimally
coupled to the curvature with coupling parameter $\xf$. Although
we could include a specific potential (quadratic or exponential),
for the scope of the present work and for simplicity we keep the
form (\ref{actioncan}) since our results can be easily generalized
to these potential-cases.  Lastly, the term $\cal{L}_\text{M}$
accounts for the matter content of the universe.

The presence of the non-minimal coupling leads to the effective
Newton's constant:
\begin{equation}
8\pi
G_{eff}=\k^{2}\left(1-\xf\k^{2}\phi^{2}\right)^{-1}\label{Geff}\,.
\end{equation}
The Friedmann equations and the evolution equation for the scalar
field are \cite{Uzan:1999ch}:
\begin{equation}
 H^{2}-\frac{\k^{2}\left(\rho_\text{M}+\rho_{\L}+\frac{1}{2}\dot{\phi}^{2}+6\xf
H\phi\dot{\phi}\right)}{3\left(1-\xf\k^{2}\phi^{2}\right)}
=0\label{eqn4}
\end{equation}
\begin{equation}
\ddot{\phi}+3H\dot{\phi}+6\xf\left(\dot{H}+2H^{2}\right)\phi=0\label{eqn5}
\end{equation}
\begin{equation}
\dot{\rho}_\text{M}+\dot{\rho}_\Lambda+3H\left(\rho_\text{M}+\rho_{\L}+p_\text{M}+p_{\L}\right)=0\label{eqn6},
\end{equation}
where $H=\dot{a}/a$ is the Hubble parameter. In these expressions,
 $p_\text{M}$ and $\rho_\text{M}$ are respectively the
pressure and density of the matter content of the universe.
Finally,  $p_\L$ and $\rho_\L$ are the corresponding components of
dark energy, which as usual is attributed to the scalar field.
Since we use the Hubble scale to define holographic dark energy,
that is we take $L=H^{-1}$, (\ref{HDEdef})  can be written as
$\rho_{\Lambda}=3(8 \pi G_{eff})^{-1}H^2$, which, due to the
effective nature of the Newton's constant (\ref{Geff}), leads to:
\begin{equation}
\rho_{\L}=\frac{3}{\k^{2}}\left(1-\xf\k^{2}\phi^{2}\right)H^{2}.\label{rhoL}
\end{equation}

We are interested in extracting power-law solutions of the
cosmological model (\ref{eqn4})-(\ref{eqn6}), in the case of a
dark-energy dominated universe ($\rho_\text{M},p_\text{M}\ll 1$).
Thus, we are looking for solutions of the form:
\begin{eqnarray}
&&a(t)=a_{0} t^{r}\nonumber\\
&&\phi(t)=\phi_{0}t^{\sff} \label{powerlaw}.
\end{eqnarray}
Insertion of these ansatzes in equations (\ref{eqn4}),(\ref{eqn5})
yields:
\begin{eqnarray}
&&\sff(\sff-1)+3r\sff+6r(2r-1)\xf =0\nonumber\\
&&\sff+12\xf r=0\label{srcan}.
\end{eqnarray}
As we can easily see, the case of conformal coupling ($\xf=1/6$)
is not interesting since it leads to the trivial case $r=\sff=0$.
Thus, for $\xf\neq1/6$ we obtain:
\begin{eqnarray}
&&r=\frac{1}{4-24\xf}\nonumber\\
&&\sff=-\frac{3\xf}{1-6\xf}\label{srcan2},
\end{eqnarray}
leading to:
\begin{eqnarray}
&&a(t)=a_{0}\;t^{\frac{1}{4-24\xf}}\nonumber\\
&&\phi(t)=\phi_{0}\;t^{-\frac{3\xf}{1-6\xf}}\label{eqn11}.
\end{eqnarray}
We can use expression (\ref{rhoL}) in order to acquire
$\rho_\L(t)$:
\begin{equation}
\rho_\L(t)=\frac{3}{\k^2}r^2\left(t^{-2}-\xf\k^2\phi_0^2
t^{2\sff-2}\right). \label{rhoLt}
\end{equation}
Substitution into (\ref{eqn6}) then straightforwardly provides
$p_{\L}$:
\begin{equation}
p_\L(t)=\frac{1}{\k^2}r\left[t^{-2}(2-3r)+\xf\k^2\phi_0^2
t^{2\sff-2}(3r+2\sff-2)\right]. \label{pLt}
\end{equation}
In expressions (\ref{rhoLt}) and (\ref{pLt}), $r$ and $\sff$ are
given by (\ref{srcan2}). Hence, we can calculate the dark energy
equation-of-state parameter $w_{\L}(t)$ as:
\begin{equation}
w_{\L}(t)=\frac{p_{\L}(t)}{\rho_{\L}(t)}=
\frac{5}{3}-8\xf\left(2+\frac{\xf\k^2\phi_0^2}{t^{-2\sff}-\xf\k^2\phi_0^2}\right)
\label{wlcan}.
\end{equation}

Relation (\ref{wlcan}) allows us to determine both the of
$w_\L$-evolution, as well as its current value $w_{\L0}$. In order
to express it in a more convenient form for comparison with
observations, we can set the current values $a_0=1$ and $t_0=1$,
and use $r\ln t=\ln a=-\ln(1+z)$ with $z$ the redshift. Therefore,
we acquire:
\begin{equation}
w_{\L}(z)=
\frac{5}{3}-8\xf\left[2+\frac{\xf\k^2\phi_0^2}{e^{-24\xf\ln(1+z)}-\xf\k^2\phi_0^2}\right]
\label{wlcan2}.
\end{equation}
Expression (\ref{wlcan2}) provides  $w_\L(z)$ in terms of the
coupling parameter $\xf$ and the amplitude $\phi_0$. We discuss
the cosmological implications for various sub-classes of the
present model, in section \ref{cosmimpl}.

\subsection{Phantom field} \label{phantom}

In this subsection we consider a phantom field with a non-minimal
coupling, that is a field with an opposite sign in the kinetic
term in the Lagrangian \cite{phant,phantBigRip}. Such models are
widely used in order to acquire $w_\L$ less than $-1$. The action
of the universe is
\begin{equation}
S=\int d^{4}x \sqrt{-g} \left[\frac{1}{2\k^{2}}
R-\frac{1}{2}\,\xs\sigma^{2} R
+\frac{1}{2}g^{\mu\nu}\p_{\mu}\sigma\p_{\nu}\sigma
+\cal{L}_\text{M}\right], \label{actionphan}
\end{equation}
and the presence of the non-minimal coupling leads to the
effective Newton's constant:
\begin{equation}
8\pi
G_{eff}=\k^{2}\left(1-\xs\k^{2}\s^{2}\right)^{-1}\label{Geff2}\,.
\end{equation}
The cosmological equations and the evolution equation for the
phantom field are \cite{phant}:
\begin{equation}
 H^{2}-\frac{\k^{2}\left(\rho_\text{M}+\rho_{\L}-\frac{1}{2}\dot{\sigma}^{2}+6\xs
H\sigma\dot{\sigma}\right)}{3\left(1-\xs\k^{2}\sigma^{2}\right)}
=0\label{eqn4b}
\end{equation}
\begin{equation}\ddot{\sigma}+3H\dot{\sigma}-6\xs\left(\dot{H}+2H^{2}\right)\sigma=0\label{eqn5b}
\end{equation}
\begin{equation}
\dot{\rho}_\text{M}+\dot{\rho}_\Lambda+3H\left(\rho_\text{M}+\rho_{\L}+p_\text{M}+p_{\L}\right)=0\label{eqn6b}.
\end{equation}
Similarly to the previous subsection, the use of the Hubble scale
in the definition of holographic dark energy, and the effective
nature of the Newton's constant (\ref{Geff2}), lead to:
\begin{equation}
\rho_{\L}=\frac{3}{\k^{2}}\left(1-\xs\k^{2}\sigma^{2}\right)H^{2}.\label{rhoL2}
\end{equation}

We examine power-law solutions of equations
(\ref{eqn4b})-(\ref{eqn6b}), in the case of a dark-energy
dominated universe ($\rho_\text{M},p_\text{M}\ll 1$). Thus, we
impose:
\begin{eqnarray}
&&a(t)=a_{0} t^{r}\nonumber\\
&&\sigma(t)=\sigma_{0}t^{\sss} \label{powerlaw2}.
\end{eqnarray}
Insertion in equations (\ref{eqn4b}),(\ref{eqn5b}) yields:
\begin{eqnarray}
&&\sss(\sss-1)+3r\sss-6r(2r-1)\xs =0\nonumber\\
&&\sss-12\xs r=0\label{srphan}.
\end{eqnarray}
As we can easily see, the case $\xs=-1/6$ leads to the trivial
case $r=\sss=0$. Thus, for $\xs\neq-1/6$ we obtain:
\begin{eqnarray}
&&r=\frac{1}{4+24\xs}\nonumber\\
&&\sss=\frac{3\xs}{1+6\xs}\label{srphan2},
\end{eqnarray}
leading to:
\begin{eqnarray}
&&a(t)=a_{0}\;t^{\frac{1}{4+24\xs}}\nonumber\\
&&\sigma(t)=\sigma_{0}\;t^{\frac{3\xs}{1+6\xs}}\label{eqn11b}.
\end{eqnarray}
Using (\ref{rhoL2}) we acquire:
\begin{equation}
\rho_\L(t)=\frac{3}{\k^2}r^2\left(t^{-2}-\xs\k^2\sigma_0^2
t^{2\sss-2}\right), \label{rhoLtb}
\end{equation}
and thus (\ref{eqn6b}) gives:
\begin{equation}
p_\L(t)=\frac{1}{\k^2}r\left[t^{-2}(2-3r)+\xs\k^2\sigma_0^2
t^{2\sss-2}(3r+2\sss-2)\right], \label{pLtb}
\end{equation}
where $r$ and $\sss$ are given by (\ref{srphan2}). We can
calculate the dark energy equation-of-state parameter $w_{\L}(t)$
as:
\begin{equation}
w_{\L}(t)=\frac{p_{\L}(t)}{\rho_{\L}(t)}=
\frac{5}{3}+8\,\xs\left(2+\frac{\xs\k^2\sigma_0^2}{t^{-2\sss}-\xs\k^2\sigma_0^2}\right)
\label{wlphan}.
\end{equation}
Finally, similarly to the previous subsection, we can express
(\ref{wlphan}) in terms of the redshift $z$ obtaining:
\begin{equation}
w_{\L}(z)=
\frac{5}{3}+8\,\xs\left[2+\frac{\xs\k^2\sigma_0^2}{e^{24\xs\ln(1+z)}-\xs\k^2\sigma_0^2}\right]
\label{wlphan2}.
\end{equation}
Relation (\ref{wlphan2}) provides $w_\L(z)$ in terms of the
coupling parameter $\xs$ and the amplitude $\sigma_0$. We examine
it in detail in section \ref{cosmimpl}.

\subsection{Quintom model} \label{phantom}

In this subsection we consider the quintom cosmological scenario
\cite{quintom}, that is we consider simultaneously a canonical and
a phantom field, both with non-minimally coupling. As we have
stated in the introduction, this combined cosmological paradigm
has been shown to be capable to describe the crossing of the
phantom divide $w_\L=-1$. The action of the model is:
\begin{eqnarray}
S=\int d^{4}x \sqrt{-g} \left[\frac{1}{2\k^{2}}
R-\frac{1}{2}\,\xf\phi^{2} R-\frac{1}{2}\,\xs\sigma^{2} R-\right.\nonumber\\
\left.
-\frac{1}{2}g^{\mu\nu}\p_{\mu}\phi\p_{\nu}\phi+\frac{1}{2}g^{\mu\nu}\p_{\mu}\sigma\p_{\nu}\sigma
+\cal{L}_\text{M}\right], \label{actionquint}
\end{eqnarray}
and the presence of the non-minimal coupling leads to the
effective Newton's constant:
\begin{equation}
8\pi
G_{eff}=\k^{2}\left[1-\k^{2}(\xf\phi^{2}+\xs\sigma^{2})\right]^{-1}\label{Geff3}\,.
\end{equation}
The cosmological equations and the evolution equation for the
canonical and phantom fields are \cite{quintom}:
\begin{equation}
H^{2}-\frac{\k^{2}\left(\rho_\text{M}+\rho_{\L}+\frac{1}{2}\dot{\phi}^{2}-\frac{1}{2}\dot{\sigma}^{2}+6\xf
H\phi\dot{\phi}+6\xs
H\sigma\dot{\sigma}\right)}{3\left[1-\k^{2}(\xf\phi^{2}+\xs\sigma^{2})\right]}
=0\label{eqn4c}
\end{equation}
\begin{equation}\ddot{\phi}+3H\dot{\phi}+6\xf\left(\dot{H}+2H^{2}\right)\phi=0\label{eqn5c}
\end{equation}
\begin{equation}\ddot{\sigma}+3H\dot{\sigma}-6\xs\left(\dot{H}+2H^{2}\right)\sigma=0\label{eqn5c2}
\end{equation}
\begin{equation}\dot{\rho}_\text{M}+\dot{\rho}_\Lambda+3H\left(\rho_\text{M}+\rho_{\L}+p_\text{M}+p_{\L}\right)=0\label{eqn6c}.
\end{equation}
As usual, the use of the Hubble scale in the definition of
holographic dark energy, and the effective nature of the Newton's
constant (\ref{Geff3}), lead to:
\begin{equation}
\rho_{\L}=\frac{3}{\k^{2}}\left[1-\k^{2}(\xf\phi^{2}+\xs\sigma^{2})\right]H^{2}.\label{rhoL3}
\end{equation}

We examine power-law solutions of equations
(\ref{eqn4c})-(\ref{eqn6c}), in the case of a dark-energy
dominated universe ($\rho_\text{M},p_\text{M}\ll 1$). Thus, we
impose:
\begin{eqnarray}
&&a(t)=a_{0} t^{r}\nonumber\\
&&\phi(t)=\phi_{0}t^{\sff}\nonumber\\
&&\sigma(t)=\sigma_{0}t^{\sss} \label{powerlaw3}.
\end{eqnarray}
Substituting into (\ref{eqn4c}),(\ref{eqn5c}),(\ref{eqn5c2}) and
requiring a solution for all times we get:
\begin{eqnarray}
&&\sff(\sff-1)+3r\sff+6r(2r-1)\xf =0\nonumber\\
&&\sss(\sss-1)+3r\sss-6r(2r-1)\xs =0\nonumber\\
&&\sff+12\xf r=0\nonumber\\
 &&\sss-12\xs r=0\label{srquint}.
\end{eqnarray}
In the present quintom scenario with both fields non-minimally
coupled, it becomes clear that the existence of non-trivial
solutions requires a relation between the couplings $\xf$ and
$\xs$.  Thus, for the physically interesting case $\xf\neq1/6$, we
obtain:
\begin{eqnarray}
&&r=\frac{1}{4-24\xf}\nonumber\\
&&\sff=-\frac{3\xf}{1-6\xf}\nonumber\\
&&\sss=-\frac{3\xf}{1-6\xf}\nonumber\\
&&\xs=-\xf \label{srquint2},
\end{eqnarray}
leading to:
\begin{eqnarray}
&&a(t)=a_{0}\;t^{\frac{1}{4-24\xf}}\nonumber\\
&&\phi(t)=\phi_{0}\;t^{-\frac{3\xf}{1-6\xf}}\nonumber\\
&&\sigma(t)=\sigma_{0}\;t^{-\frac{3\xf}{1-6\xf}}\label{eqn11c}.
\end{eqnarray}
Note also that we could equivalently express solutions
(\ref{srquint2}) in terms of $\xs$. Finally, as expected, the
choice $\xf=1/6$ gives $-1/6<\xs$, which is also non-physical.

Using (\ref{rhoL3}) we obtain:
\begin{equation}
\rho_\L(t)=\frac{3}{\k^2}r^2\left(t^{-2}-\xf\k^2\phi_0^2
t^{2\sff-2}-\xs\k^2\sigma_0^2 t^{2\sss-2}\right), \label{rhoLtc}
\end{equation}
and thus (\ref{eqn6c}) gives:
\begin{eqnarray}
p_\L(t)=&&\frac{1}{\k^2}r\left[t^{-2}(2-3r)+\xf\k^2\phi_0^2
t^{2\sff-2}(3r+2\sff-2)+\right.\nonumber\\
&&\left.+\xs\k^2\sigma_0^2 t^{2\sss-2}(3r+2\sss-2)\right],
\label{pLtc}
\end{eqnarray}
where $r$, $\sff$, $\sss$ and $\xs$ are given by (\ref{srquint2}).
Thus, we can calculate the dark energy equation-of-state parameter
$w_{\L}(t)$ as:
\begin{equation}
w_{\L}(t)=\frac{p_{\L}(t)}{\rho_{\L}(t)}=
\frac{5}{3}-16\xf+8\left[\frac{\k^2\left(\xf^2\phi_0^2t^{2\sff}-\xs^2\sigma_0^2t^{2\sss}\right)}
{\k^2\left(\xf\phi_0^2t^{2\sff}+\xs\sigma_0^2t^{2\sss}\right)-1}\right]
\label{wlquint}.
\end{equation}
Finally, expressing (\ref{wlquint}) in terms of the redshift $z$
we obtain: {\small{
\begin{eqnarray}
&&w_{\L}(z)=
\frac{5}{3}-16\xf+\nonumber\\
&&+8\left\{\frac{\k^2\left[\xf^2\phi_0^2e^{24\xf\ln(1+z)}-\xs^2\sigma_0^2e^{-24\xs\ln(1+z)}\right]}
{\k^2\left[\xf\phi_0^2e^{24\xf\ln(1+z)}+\xs\sigma_0^2e^{-24\xs\ln(1+z)}\right]-1}\right\}\
\ \ \  \label{wlquint2}
\end{eqnarray}}}
Relation (\ref{wlquint2}) provides $w_\L(z)$ in terms of the
coupling parameters $\xf$, $\xs$ and the amplitudes $\phi_0$,
$\sigma_0$. Note that (\ref{wlquint2}) corresponds to the quintom
scenario, and thus expressions (\ref{srquint2}) are embedded in
it. Therefore, one cannot simply set some parameters to zero in
order to obtain the simple canonical or simple phantom cases, but
he has to solve the problem from the beginning with only one
field, that is the procedure we followed in the previous
subsections. In the next section we analyze the cosmological
implications of the quintom model.

\section{Cosmological implications} \label{cosmimpl}

In the previous subsections we have obtained the equation-of-state
parameter of dark energy $w_\L(z)$, in terms of the coupling
parameters $\xf$, $\xs$ and the amplitudes $\phi_0$, $\sigma_0$.
In the present section we investigate the cosmological
implications for each case.

\subsection{Canonical field} \label{cosmimplcan}

\begin{figure}[ht]
\begin{center}
\mbox{\epsfig{figure=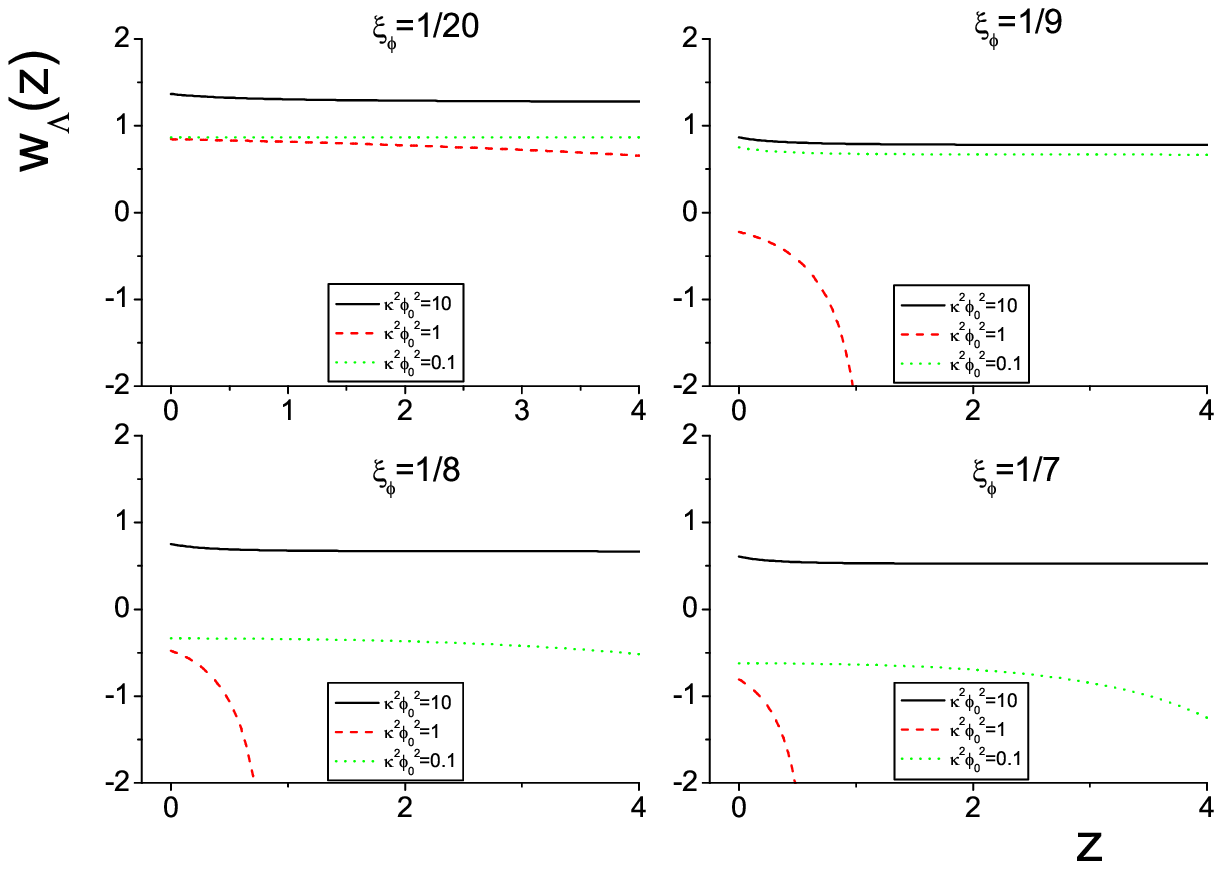,width=8.7cm,angle=0}}
\caption{(Color online) {\it  $w_\L(z)$ vs $z$ in the canonical
field case, for $\xf=1/20$, $\xf=1/9$, $\xf=1/8$, $\xf=1/7$, where
in each case the combination $\k^2\phi_0^2$ is taken equal to
$10$, $1$, $0.1$ respectively. The divergence of $w_\L(z)$ is a
direct consequence of the singularity of (\ref{wlcan2}), and thus
the corresponding combinations of $\xf$ and $\k^2\phi_0^2$ must be
excluded.}} \label{canonicalfig}
\end{center}
\end{figure}

In the case of a simple canonical field, non-minimally coupled to
gravity, $w_\L(z)$ is given by relation (\ref{wlcan2}). In
fig.~\ref{canonicalfig} we depict $w_\L(z)$ for four different
values of the coupling $\xf$ and for three different values of the
combination $\k^2\phi_0^2$. Note that the physical requirement of
an expanding universe, results to an upper limit for $\xf$, namely
$\xf<1/6$, as can be seen in the first relation (\ref{eqn11}). In
addition, we mention that in general $\xf$ could be also negative,
but since it leads to non-physical behavior of $w_\L(z)$  we
neglect this case in this subsection.

As we observe, the value of $w_\L(z)$ at $z=0$, that is its
current value $w_{\L0}$, decreases as $\xf$ increases, while its
dependence on $\k^2\phi_0^2$ is non-monotonic. However, in this
simple canonical field case $w_{\L0}$ is always greater than $-1$,
independently of the values of $\xf$ and $\k^2\phi_0^2$. This was
expected since this case is well known to be insufficient to
describe the crossing of the phantom divide $w_\L=-1$ from above
\cite{quint}.

Secondly, we can see that for not so small $\xf$, and for
$\k^2\phi_0^2$ of the order of 1, we obtain a divergence of
$w_\L(z)$. This behavior is a clear prediction of relation
(\ref{wlcan2}), since it possesses a singularity at:
\begin{equation}
z_{s}=-1+\left(\xf\k^2\phi_0^2\right)^{-\frac{1}{24\xf}}
\label{singcan}.
\end{equation}
Therefore, the combinations of $\xf$ and $\k^2\phi_0^2$ that
satisfy this transcendental equation giving a positive $z_s$, must
be excluded. However, focusing on the future instead of the past,
this behavior of $w_{\L}$ has a very important cosmological
implication. Using directly the form (\ref{wlcan}), which allows
us to investigate the future evolution, we conclude that there are
some combinations of $\xf$ and $\k^2\phi_0^2$ that lead to a
future divergence of $w_\L$. Thus, the non-minimally coupled
canonical field model of holographic dark energy predicts a
cosmological singularity
 at a future time $t_{CS}$, for
combinations of $\xf$ and $\k^2\phi_0^2$ that satisfy:
\begin{equation}
t_{CS}= \left(\xf\k^2\phi_0^2\right)^{\frac{1}{6\xf}-1}>1
\label{Bigripcan},
\end{equation}
and since $\xf<1/6$, the $w_{\L}$-divergence realization condition
reads simply:
\begin{equation}
\xf\k^2\phi_0^2>1 \label{Bigripcan2}.
\end{equation}
Fortunately, this condition leads to a negative effective Newton's
constant in (\ref{Geff}), and thus the corresponding parameter
combinations must be excluded, leaving the model free of a future
$w_{\L}$-divergence. In any case, we have to mention that in the
model at hand the $w_{\L}$-divergence at $t_{CS}$ is not
accompanied by a divergence in the scale factor, in its
time-derivative and in the dark energy density and pressure. Thus,
technically, it does not correspond to the
 Big Rip of the literature \cite{phantBigRip,BigRip},
 but rather to some new singularity family.

For reasons of completeness we present explicitly the behavior of
$w_\L(z)$ for $\k^2\phi_0^2\ll1$, that is for very small current
value of the scalar field. Specifically, we find that the present
value $w_{\L0}$ is:
\begin{equation}
\left. w_{\L0} \right|_{\k^2\phi_0^2\ll1}\simeq
\frac{5}{3}-16\xf>-1 \label{f0small},
\end{equation}
with the last inequality arising from the upper bound of
$\xf<1/6$.

Finally, we mention that the model at hand should receive
additional constraints through the observations of the time
variation of gravitational constant \cite{G4com}. In particular,
differentiating (\ref{Geff}) with respect to $t$ and setting
$t_0=1$ for the present time, we acquire:
\begin{equation}
\left.\frac{\dot{G}}{G}\right|_{0}=
-\frac{6\xf^{2}\k^{2}\phi^{2}_{0}}{\left(1-6\xf\right)\left(1-\xf\k^2\phi^{2}_{0}\right)}
,\label{Hvarcan}
\end{equation}
where we have also used (\ref{srcan2}) and  (\ref{eqn11}). This
combination must be less than $4\%$ \cite{G4com}.

\subsection{Phantom field} \label{cosmimplphan}

\begin{figure}[ht]
\begin{center}
\mbox{\epsfig{figure=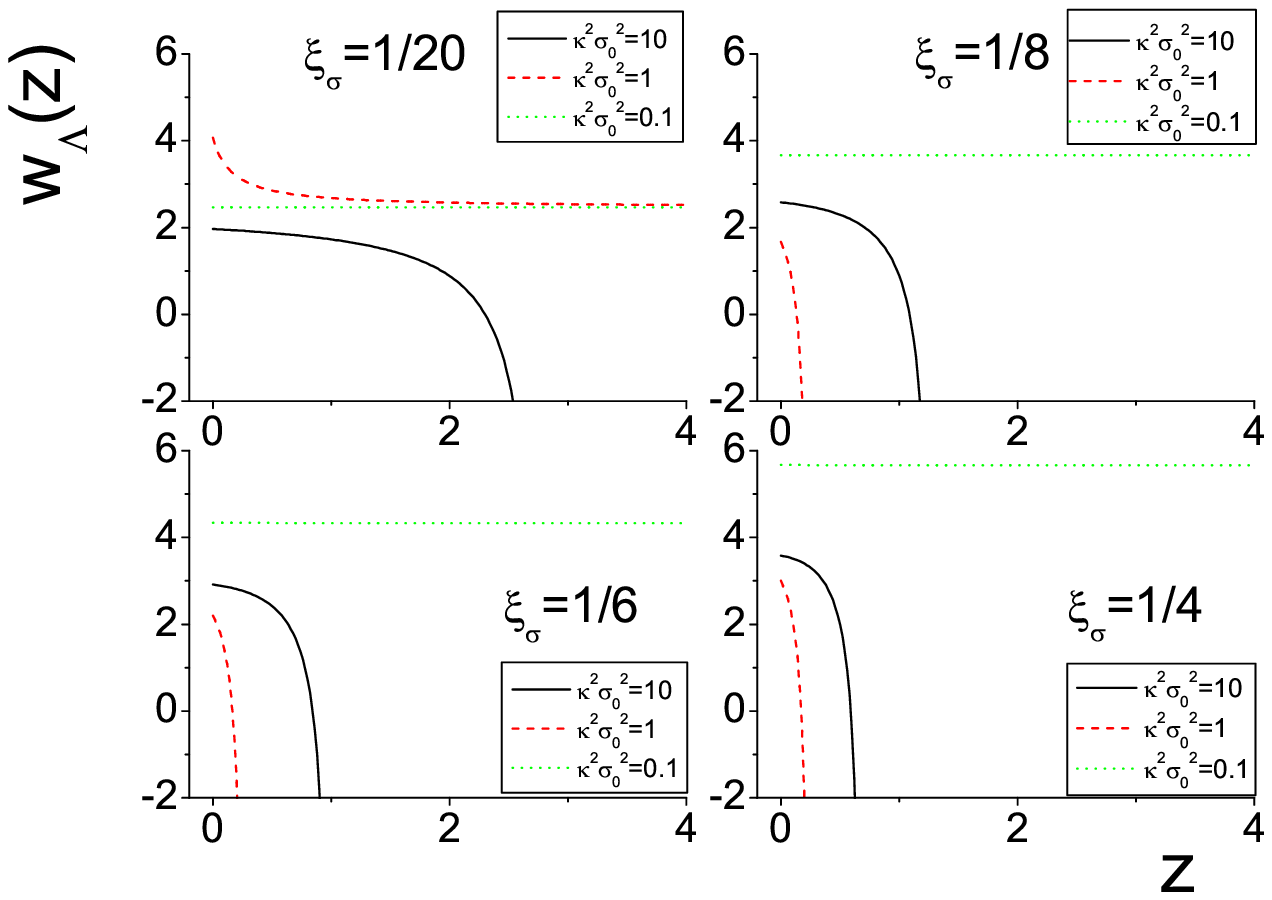,width=8.7cm,angle=0}}
\caption{(Color online) {\it  $w_\L(z)$ vs $z$ in the phantom
field case, for $\xs=1/20$, $\xs=1/8$, $\xs=1/6$, $\xs=1/4$, where
in each case the combination $\k^2\s_0^2$ is taken equal to $10$,
$1$, $0.1$ respectively. The divergence of $w_\L(z)$ is a direct
consequence of the singularity of (\ref{wlphan2}), and thus the
corresponding combinations of $\xs$ and $\k^2\s_0^2$ must be
excluded.}} \label{phantomfig}
\end{center}
\end{figure}

In the case of a phantom field, non-minimally coupled to gravity,
$w_\L(z)$ is given by relation (\ref{wlphan2}). In
fig.~\ref{phantomfig} we depict $w_\L(z)$ for four different
values of the coupling $\xs$ and for three different values of the
combination $\k^2\s_0^2$. Note that in this case the physical
requirement of an expanding universe, results to a lower limit for
$\xs$, namely $-1/6<\xs$, as it is implied by the first relation
(\ref{eqn11b}).

As we can see, the value of $w_{\L0}$ is now a non-monotonic
function of $\xs$ and $\k^2\s_0^2$. Furthermore, we observe that
for some particular combinations of $\xs$ and $\k^2\s_0^2$, as a
consequence of the singularity of (\ref{wlphan2}),  there is a
divergence of $w_\L(z)$  at:
\begin{equation}
z_{s}=-1+\left(\xs\k^2\s_0^2\right)^{\frac{1}{24\xs}}
\label{singphan}.
\end{equation}
Thus, the combinations of $\xs$ and $\k^2\s_0^2$ that satisfy this
transcendental equation giving a positive $z_s$, must be excluded.
Note that in the case of negative $\xs$,  condition
(\ref{singphan}) cannot be satisfied and thus this solution
sub-class is free of a past divergence.

Similarly to the previous subsection, using the form
(\ref{wlphan}), which allows us to investigate the future
evolution, we conclude that there are some combinations of $\xs$
and $\k^2\s_0^2$ that lead to a future divergence of $w_\L$.
Therefore, the non-minimally coupled phantom field model of
holographic dark energy predicts a a cosmological singularity
 at $t_{CS}$, for combinations of $\xs$
and $\k^2\s_0^2$ that satisfy:
\begin{equation}
t_{CS}=\left(\xs\k^2\s_0^2\right)^{-\frac{1}{6\xs}-1}>1
\label{Bigripphan}.
\end{equation}
Note that in the case of negative $\xs$, this condition cannot be
satisfied and thus this solution sub-class is free of a future
cosmological singularity. For a positive $\xs$ the
$w_{\L}$-divergence realization condition is simply
\begin{equation}
\xs\k^2\s_0^2<1 \label{Bigripcanbbbc},
\end{equation}
which  does not bring any positivity problems in the effective
Newton's constant in (\ref{Geff2}). Thus, the corresponding
parameter combinations cannot be excluded, and the model, for a
positive $\xs$, clearly predicts a future cosmological
singularity.

\begin{figure}[ht]
\begin{center}
\mbox{\epsfig{figure=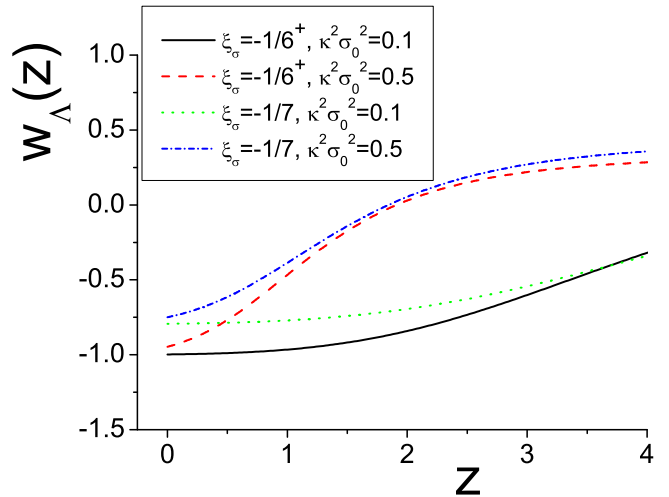,width=8.7cm,angle=0}}
\caption{(Color online) {\it  $w_\L(z)$ vs $z$ in the phantom
field case, for four different parameter choices with negative
values of $\xs$. The index $+$ in $-1/6^+$ marks the limit
$\xs\rightarrow-1/6$ from above.}} \label{phantomfig2}
\end{center}
\end{figure}

In the case at hand we can see that $w_{\L0}$ is always greater
than $-1$, independently of the values of $\xs$ and $\k^2\s_0^2$,
which is not what is expected for a phantom field. This behavior
is a clear result of the non-minimal coupling in the holographic
dark energy framework. However, contrary to the canonical field
case where negative values of the coupling lead to non-physical
behavior ($w_{\L0}>1$), in this phantom field scenario such a
choice leads to interesting cosmological implications. In fig.
\ref{phantomfig2} we depict $w_\L(z)$ for four different parameter
choices with negative values of $\xs$.
As we observe, negative values of the coupling produce
cosmological behaviors with decreasing $w_\L(z)$ and $w_{\L0}$
very close to $-1$. Furthermore, for $\xs<-1/7$ we obtain a
$w_{\L0}$ inside the observational limits  \cite{1,2,3,4},
although we cannot acquire a clear phantom divide crossing. It is
known that under specific potential choices, a non-minimally
coupled phantom scenario can achieve the $-1$-crossing
\cite{Carvalho:2004ty}. It seems that the holographic dark energy
framework does not allow for such a behavior.

Additionally, taking the limit $\k^2\s_0^2\ll1$  we find that:
\begin{equation}
\left. w_{\L0} \right|_{\k^2\s_0^2\ll1}\simeq \frac{5}{3}+16\xs>-1
\label{s0small},
\end{equation}
with the last inequality arising from the lower bound of
$-1/6<\xs$.

 Finally, the present scenario should also receive
additional constraints through the observations of the time
variation of gravitational constant \cite{G4com}. In particular,
we acquire:
\begin{equation}
\left.\frac{\dot{G}}{G}\right|_{0}=
\frac{6\xs^{2}\k^{2}\s^{2}_{0}}{\left(1+6\xs\right)\left(1-\xs\k^2\s^{2}_{0}\right)}
,\label{Hvarphan}
\end{equation}
where we have also used (\ref{srphan2}) and  (\ref{eqn11b}), and
thus this combination must be less than $4\%$ \cite{G4com}.

\subsection{Quintom model} \label{cosmimplquint}

\begin{figure}[ht]
\begin{center}
\mbox{\epsfig{figure=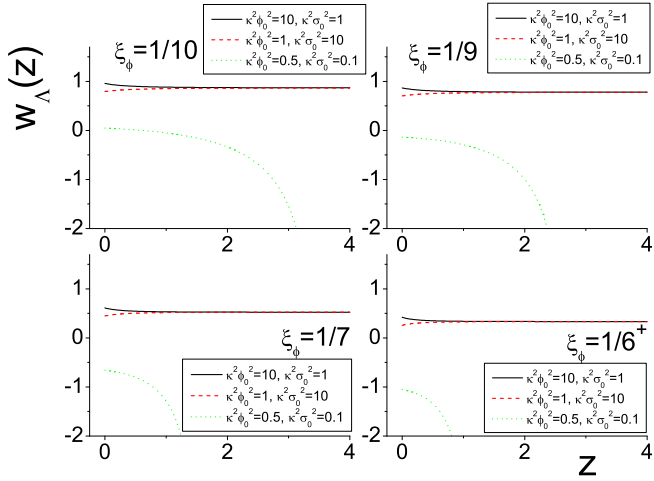,width=8.7cm,angle=0}}
\caption{(Color online) {\it  $w_\L(z)$ vs $z$ in the combined
quintom scenario, for $\xf=1/20$, $\xf=1/9$, $\xf=1/8$, $\xf=1/7$,
where in each case the combinations $\k^2\phi_0^2$ and
$\k^2\s_0^2$ are shown in the insets. The divergence of $w_\L(z)$
is a direct consequence of the singularity of (\ref{wlquint2}),
and thus the corresponding combinations of $\xf$, $\k^2\phi_0^2$
and $\k^2\s_0^2$ must be excluded.}} \label{quintomfig}
\end{center}
\end{figure}

In the case of the combined quintom model, that is when both the
canonical and phantom fields are considered to be non-minimally
coupled to gravity simultaneously, $w_\L(z)$ is given by relation
(\ref{wlquint2}). In fig.~\ref{quintomfig} we depict $w_\L(z)$ for
four different values of the coupling $\xf$ and for three
different combinations $\k^2\phi_0^2$ and $\k^2\s_0^2$. Note that
in this case the physical requirement of an expanding universe,
results to an upper limit for $\xf$, namely $\xf<1/6$, as it is
implied by the first relation (\ref{eqn11c}).
The value of $w_{\L0}$ is a monotonic function of $\xf$. As in the
previous cases, for some particular combinations of $\xf$,
$\k^2\phi_0^2$ and $\k^2\s_0^2$, as a consequence of
(\ref{wlquint2}), there is a singularity of $w_\L(z)$ at a
specific $z_{s}$. The form of the denominator of (\ref{wlquint2})
does not allow for an explicit expression of $z_{s}$, but
numerical investigation provides the specific excluded parameter
values.

Similarly to the previous subsections, there are some combinations
of $\xf$, $\k^2\phi_0^2$ and $\k^2\s_0^2$ that lead to a future
divergence of $w_\L$. Thus, the non-minimally coupled quintom
model of holographic dark energy predicts a future cosmological
singularity, for parameter combinations that make (\ref{wlquint})
diverge for $t_{CS}>1$, that is in the future. We mention that the
transcendental form of the denominator forbids the extraction of
an explicit relation for $t_{CS}$, but the corresponding values
can be provided by simple numerical calculations.

\begin{figure}[ht]
\begin{center}
\mbox{\epsfig{figure=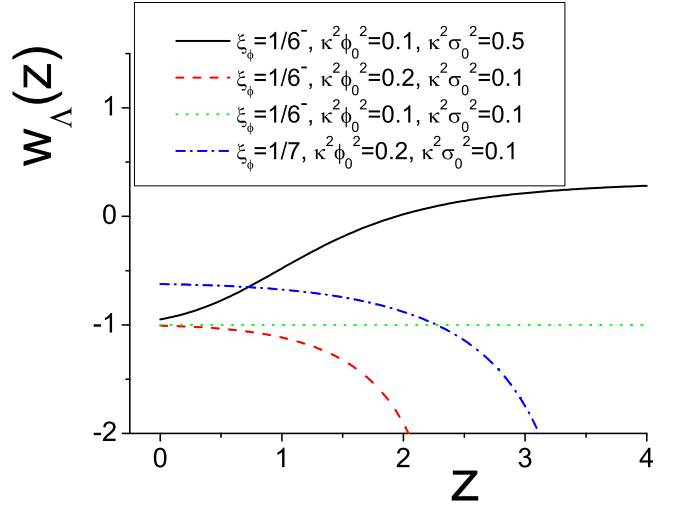,width=8.7cm,angle=0}}
\caption{(Color online) {\it  $w_\L(z)$ vs $z$ in the quintom
case, for four combinations of $\xf$, $\k^2\phi_0^2$ and
$\k^2\s_0^2$ shown in the inset. The index $-$ in $1/6^-$ marks
the limit $\xf\rightarrow1/6$ from below.}} \label{quintomfig2}
\end{center}
\end{figure}

As we observe in fig.~\ref{quintomfig}, $w_{\L0}$ is greater than
$-1$, independently of the values of $\xf$, $\k^2\phi_0^2$  and
$\k^2\s_0^2$. However, for a class of parameter combinations we
obtain cosmological evolutions in agreement with observations. In
fig. \ref{quintomfig2} we depict $w_\L(z)$ for four such
combinations of $\xf$, $\k^2\phi_0^2$ and $\k^2\s_0^2$.
As we can see, we can obtain a decreasing form of $w_\L$ with its
current values inside the observational limits \cite{1,2,3,4}. It
is interesting that we cannot acquire a clear $-1$-crossing, which
was the basic motive of the construction of quintom scenario
\cite{quintom}. It seems that the holographic dark energy
framework refutes such an eventuality. Furthermore, we mention
that in the case where $\k^2\phi_0^2=\k^2\s_0^2$ the effects of
the canonical and phantom fields cancel each other, as expected by
relation (\ref{wlquint2}), and the dark energy in the model at
hand behaves like a cosmological constant (dotted curve of fig.
\ref{quintomfig2}). Lastly, numerical investigations show that the
parameter subspace that leads to $w_{\L0}\approx-1$, cannot lead
to a future cosmological singularity, which is also an advantage
of the model.

Taking the limit $\k^2\phi_0^2\s_0^2\ll1$ we find that:
\begin{equation}
\left. w_{\L0} \right|_{\k^2\phi_0^2\s_0^2\ll1}\simeq
\frac{5}{3}-16\xf>-1 \label{f0small},
\end{equation}
with the last inequality arising from the upper bound of
$\xf<1/6$.  Finally, we close this subsection with the external
constraints to the model by the observations of the time variation
of gravitational constant \cite{G4com}. In particular, we acquire:
\begin{equation}
\left.\frac{\dot{G}}{G}\right|_{0}=-\left[1-\k^{2}(\xf\phi_0^{2}+\xs\sigma_0^{2})\right]^{-1}\left[\frac{6\xf^{2}\k^{2}\phi^{2}_{0}}{\left(1-6\xf\right)}
-\frac{6\xs^{2}\k^{2}\s^{2}_{0}}{\left(1+6\xs\right)}\right]
,\label{Hvarphan}
\end{equation}
where we have also used (\ref{srquint2}) and  (\ref{eqn11c}), and
thus this combination must be less than $4\%$ \cite{G4com}.

\section{Conclusions} \label{conclusions}

In this work we construct various field models of dark energy,
such as simple canonical and phantom fields, and their
simultaneous consideration into a combined model called quintom.
All fields are non-minimally coupled to gravity through extra
terms in the action, and the investigation has been performed in
the framework of holographic dark energy. In each case we extract
$w_\L(z)$, that is the dark energy equation-of-state parameter, as
a function of the redshift and using as parameters the couplings
and the amplitudes of the fields, and we analyze it in order to
obtain its cosmological implications. In particular we examine the
present value $w_{\L0}$, the crossing through the phantom divide
$-1$, and we extract the conditions for a future cosmological
singularity.

For the simple canonical field we find that $w_{\L0}$ cannot be
less than $-1$, thus this model cannot describe the transition
through the phantom divide. In addition, we give the parameter
subspace that has to be excluded since it leads to a singular
behavior in the past. Furthermore, we find that for a specific
parameter subspace the universe will result in a future
cosmological singularity, and we extract a specific relation for
the time that it is going to be realized. Fortunately, the
physical requirement for a positive effective Newton's constant
makes the model free of such a singularity. Finally, we give a
constraint for the model parameters in order for the time
variation of the gravitational constant to be consistent with
observations.

For the simple phantom field we provide the parameter subspace
that has to be excluded in order to acquire a regular evolution in
the past. We extract the conditions and the time of a future
cosmological singularity. For the case of negative couplings we
find a decreasing $w_\L$ with a current value inside the
observational limits, in agreement with cosmological observations.
The fact that  $w_\L$ lies above the phantom divide is a clear
result of the non-minimal coupling in the holographic dark energy
framework. Furthermore, in the single phantom field case, the
future cosmological singularity cannot be excluded. Lastly, we
present the constraints to the model by the time variation of the
gravitational constant.

For the quintom model, that is the combined case of both canonical
and phantom fields, we give the conditions for physical
evolutions, that is without divergencies in the past, and we
provide the requirements for a future cosmological singularity. We
find that a clear crossing of the phantom divide cannot be
obtained, in contrast to what is expected for a quintom scenario.
It seems that the holographic dark energy framework refutes such a
behavior. However, we do obtain a decreasing $w_\L$ with a current
value inside the observational limits. In addition, these
 solutions do not possess a future cosmological singularity and these
features make them a good candidate for the description of dark
energy. Finally we provide the parameter constraints in order for
the model to be consistent with the observed time variation of the
gravitational constant.\\

\paragraph*{{\bf{Acknowledgements:}}}
The authors would like to thank an anonymous referee for crucial
remarks and advices. E. N. Saridakis wishes to thank Institut de
Physique Th\'eorique, CEA, for the hospitality during the
preparation of the present work.

\end{document}